\documentclass[prc,twocolumn,superscriptaddress,superscriptfootnote]{revtex4}
\usepackage{amssymb}

\usepackage{amsmath,amsfonts,amssymb,epsfig,color}

\begin{document}
\title{Structure of the doublet bands in doubly odd nuclei: The case of $^{128}Cs$}
\author{H. G. Ganev}
\affiliation{Institute of Nuclear Research and Nuclear Energy,
Bulgarian Academy of Sciences,\\ Sofia 1784, Bulgaria}
\author{S. Brant}
\affiliation{Department of Physics, Faculty of Science, University
of Zagreb, 10000 Zagreb, Croatia}

\setcounter{MaxMatrixCols}{10}

\begin{abstract}
The structure of the $\Delta J = 1$ doublet bands in $^{128}Cs$ is
investigated within the framework of the Interacting Vector Boson
Fermion Model (IVBFM). A new, purely collective interpretation of
these bands is given on the basis of the used boson-fermion
dynamical symmetry of the model. The energy levels of the doublet
bands as well as the absolute $B(E2)$ and $B(M1)$ transition
probabilities between the states of both yrast and yrare bands are
described quite well. The observed odd-even staggering of both
$B(M1)$ and $B(E2)$ values is reproduced by the introduction of an
appropriate interaction term of quadrupole type, which produces such
a staggering effect in the transition strengths. The calculations
show that the appearance of doublet bands in certain odd-odd nuclei
could be a consequence of the realization of a larger dynamical
symmetry based on the non-compact supersymmetry group $OSp(2\Omega
/12, R)$.

\end{abstract}
\maketitle PACS {21.10.Re,23.20.Lv,21.60.Ev,27.60.+j}

\section{Introduction}

One of the most intriguing phenomena which has attracted significant
attention and discussed intensively in the last decade is the
appearance of the nearly degenerate $\Delta J = 1$ doublet bands
with the same spins and parities in odd-odd $N=75$ and $N=73$
isotones in the $A\sim 130$ region. A large number of experimental
data \cite{a1}-\cite{a6},\cite{cphcm},\cite{a8} have been
accumulated in this mass region which showed that the yrast and
yrare states are built on the $\pi h_{11/2} \otimes \nu h_{11/2}$
configuration. Pairs of bands have been found also in the $A \sim
105$ and $A \sim 190$ mass regions. Initially, these $\Delta J = 1$
doublet bands had been interpreted as a manifestation of "chirality"
in the sense of the angular momentum coupling \cite{FM}. Several
theoretical models have been applied in a number of articles, like
the tilted axis cranking (TAC) model
\cite{a1}-\cite{a2},\cite{a8},\cite{tac3}, the core-quasiparticle
coupling model (CQPCM) \cite{cqpcm2}, the particle-rotor model (PRM)
\cite{a4}-\cite{a5},\cite{prm3}, two quasiparticle + triaxial rotor
model (TQPTR) \cite{tqptrm}, core-particle-hole coupling model
(CPHCM) \cite{cphcm}. All these models have one assumption in
common, they suppose a rigid triaxial core and hence support the
interpretation of the doublet bands of chiral structure. On the
contrary, all odd-odd nuclei in which twin bands have been observed
have a different characteristics in common, they are in regions
where even-even nuclei are $\gamma$-soft. Their potential energy
surface is rather flat in the $\gamma$-direction and the couplings
with other core structures, not only the ground state band, are
significant. Nevertheless, it was shown in \cite{soft} that the
odd-odd nuclei with soft cores have chiral properties similar to
those with the rigid core structures. Although the odd-odd nuclei in
$A\sim 130$ region should properly be described by soft cores the
rigidity does not seem to be decisive for chirality \cite{cphcm}.

Many of the recent experiments and theoretical analysis do not
support completely the chiral interpretation
\cite{contra1}-\cite{contra4}. In particular, in an ideal situation,
i.e. orthogonal angular momentum vectors and stable triaxial nuclear
shape, states with the same spin should be observed close in
excitation energy. In fact, the attainment of such near-degeneracy
is one of the key characteristics of chirality. This feature has not
been observed in any of the chiral structures identified to date.
Moreover, states with different structure in two nonchiral bands can
also accidentally be close in excitation energy. Thus, one of the
important tests of chirality is that the partner states in the two
bands should also have similar physical properties, such as moment
of inertia, quasiparticle alignments, transition quadrupole moments,
and the related $B(E2)$ values for intraband $E2$ transitions. Some
experimental studies have shown that the two bands have different
shapes due to the different kinematical moments of inertia, which
suggest a shape coexistence (triaxial and axial shapes). This is an
interesting observation since the quantal nature of chirality
automatically demands that a chiral partner band should have
identical properties to the yrast triaxial rotational band.
Similarly, it was also found for some of the proposed chiral nuclei
that the experimental data for the behavior of other observables
(equal $E2$ transitions increasing with spin, staggering behavior of
the $M1$ values, the smoothness of the signature $S(J)$, etc.) do
not support such a chiral structure \cite{contra1}-\cite{contra4}.
These results demand a deeper and more detailed discussion of our
understanding of the origin of doublet bands. Although the odd-odd
nuclei in the considered mass region do not satisfy all the
requirements for the existence of chirality, they can approach some
of them, or at least retain some fingerprints of chirality. In this
respect it is appropriated to consider the observation of nearly
degenerate doublet bands exhibiting some of the chiral features as
manifestation of the (weak) chiral symmetry breaking phenomenon
\cite{a6},\cite{breaking}.

The fact that two bands of the same parity have levels of the same
spin close in excitation energy is not a very strong argument to
claim that they are chiral bands. In order to establish their chiral
structure it is crucial to determine the $B(E2)$ and $B(M1)$ values.
In this respect, the lifetime measurements are essential for
extracting the absolute $B(E2)$ and $B(M1)$ transition
probabilities, which are critical experimental observables in
addition to the energy levels. In a number of papers (e.g.
\cite{M1}) the observation of $B(M1)$ staggering was suggested as
the main fingerprint for the identification of the chiral doublet
bands. The strong staggering is considered as manifestation of the
static chirality, whereas the weak staggering was interpreted as a
chiral vibration \cite{chivib}. Characteristic properties of the
chiral bands are closely connected with the triaxiality (rigid or
soft) and the deviation from maximal triaxiality causes fast
splitting of the partner bands \cite{soft} and the vanishing of the
$B(M1)$ (and $B(E2)$) staggering \cite{M1stag}.

Within the framework of the pair truncated shell model (PTSM) it was
pointed out that the band structure of the doublet bands can be
explained by the chopsticks-like motion of two angular momenta of
the odd neutron and the odd proton \cite{ptsm}.  It was found that
the level scheme of $\Delta J = 1$ doublet bands does not arise from
the chiral structure, but from different angular momentum
configurations of the unpaired neutron and unpaired proton in the
$0h_{11/2}$ orbitals, weakly coupled with the collective excitations
of the even-even core. The same interpretation was given also in the
quadrupole coupling model (QCM) \cite{qcm1},\cite{qcm2}.

An alternative interpretation is based on the Interacting boson
fermion-fermion model (IBFFM) \cite{IBFFM}, where the energy
degeneracy is obtained but a different nature is attributed to the
two bands. A detailed analysis of the wave functions in IBFFM showed
as well that the presence of configurations with the angular momenta
of the proton, neutron and core in the chirality favorable, almost
orthogonal geometry, is substantial but far from being dominant. The
large fluctuations of the deformation parameters $\beta$ and
$\gamma$ around the triaxial equilibrium shape enhance the content
of achiral configurations in the wave functions. The composition of
the yrast band, in terms of contributions from core states, shows
that the yrast band is basically built on the ground-state band of
the even-even core. With increasing spin the admixture of the
$\gamma-$band of the core becomes more pronounced. The side band
wave functions contain large components of the $\gamma-$band and
with increasing spin, of higher-lying collective structures of the
core, which near the band crossing become dominant. Thus, according
to IBFFM the existence of twin bands should be attributed to a weak
dynamic (fluctuation dominated) chirality.

The above variety of models and approaches dealing with the
description of the doublet bands in odd-odd nuclei reveals the
complexity of the chiral rotation and  motivated us to consider
their properties in the framework of the boson-fermion extension
\cite{OSE} of the symplectic Interacting Vector Boson Model (IVBM),
for which we will use the term Interacting Vector Boson Fermion
Model (IVBFM). In \cite{DB} the investigation of the doublet bands
in some doubly odd nuclei from $A\sim130$ region was presented. A
good agreement between experiment and theoretical predictions for
the energy levels of these bands as well as in-band $B(E2)$ and
$B(M1)$ transition probabilities is obtained. With the present work
we exploit further the new dynamical symmetry \cite{OSE},\cite{DB}
of IVBFM for the analysis of the structure of the $\pi h_{11/2}
\otimes \nu h_{11/2}$ positive-parity doublet bands in $^{128}Cs$.
Recently, lifetime measurements in $^{128}Cs$ were performed to
extract the absolute transition probabilities $B(M1)$ and $B(E2)$ to
identify candidate chiral doublet bands \cite{Cs128}. The partner
bands in $^{128}Cs$ with similar $B(M1)$ and $B(E2)$ transitions and
strong $B(M1)$ staggering were observed and regarded as the best
known example revealing the chiral symmetry breaking phenomenon
\cite{a6},\cite{breaking}.  A systematic study of doublet bands in
the nearby odd-odd $^{128-134}Cs$ isotopes have been done in
\cite{a6}. Attempting to search for the chiral doublet bands in
$^{126}Cs$, high-spin states of $^{126}Cs$ were investigated in
\cite{Cs126} and candidate chiral doublet bands in $^{126}Cs$ were
proposed. Based on a systematic comparison with the neighboring
odd-odd Cs isotopes, a pair of chiral doublet bands in $^{122}Cs$
are proposed in \cite{Cs122}.

The spectrum of the positive-parity states in $^{128}Cs$ considered
in this paper is based on the odd proton and odd neutron that occupy
the same single particle level $h_{11/2}$. The theoretical
description of the doubly odd nuclei under consideration is fully
consistent and starts with the calculation of theirs even-even and
odd-even neighbors. We consider the simplest physical picture in
which two particles occupying the same single particle level $j$ are
coupled to an even-even core nucleus whose states belong to an
$Sp(12,R)$ irreducible representation. Within the framework of IVBFM
a purely collective structure of the doublet bands is obtained. To
describe the structure of odd-odd nuclei, first a description of the
appropriate even-even cores should be obtained.

\section{The even-even core nucleus}

The algebraic structure of IVBM is realized in terms of creation and
annihilation operators of two kinds of vector bosons
$u_{m}^{+}(\alpha )$, $u_{m}(\alpha )$ ($m=0,\pm 1$), which differ
in an additional quantum number $\alpha=\pm1/2-$the projection of
the $T-$spin (an analogue to the $F-$spin of IBM-2). One might
consider these two bosons just as building blocks or "quarks" of
elementary excitations (phonons) rather than real fermion pairs,
which generate a given type of symmetry. In this regard, the $s$ and
$d$ bosons of the IBM can be considered as bound states of
elementary excitations generated by two vector bosons. Thus, we
assume that it is the type of symmetry generated by the bosons which
is of importance for the description of the collective motions in
nuclei.

All bilinear combinations of the creation and annihilation operators
of the two vector bosons generate the boson representations of the
non-compact symplectic group $ Sp^{B}(12,R)$ \cite{GGG}:
\begin{eqnarray}
F_{M}^{L}(\alpha ,\beta ) = {\sum }_{k,m}C_{1k1m}^{LM}u_{k}^{+}(
\alpha )u_{m}^{+}(\beta ), \label{Fs} \\
G_{M}^{L}(\alpha ,\beta ) ={\sum }_{k,m}C_{1k1m}^{LM}u_{k}(\alpha
)u_{m}(\beta ), \label{Gs} \\
A_{M}^{L}(\alpha, \beta )={\sum }_{k,m}C_{1k1m}^{LM}u_{k}^{+}(\alpha
)u_{m}(\beta ),  \label{numgen}
\end{eqnarray}
where $C_{1k1m}^{LM}$, which are the usual Clebsch-Gordan
coefficients for $L=0,1,2$ and $M=-L,-L+1,...L$, define the
transformation properties of (\ref{Fs}),(\ref{Gs}) and
(\ref{numgen}) under rotations. Being a noncompact group, the
unitary representations of $Sp^{B}(12,R)$ are of infinite dimension,
which makes it impossible to diagonalize the most general
Hamiltonian. When reduced to the group $U^{B}(6)$, each irrep of the
group $Sp^{B}(12,R)$ decomposes into irreps of the subgroup
characterized by the partitions \cite{GGG}: $\lbrack
N,{0}^5\rbrack_{6}\equiv \lbrack N]_{6}$, where $N=0,2,4,\ldots$
(even irrep) or $N=1,3,5,\ldots$ (odd irrep). The subspaces
$[N]_{6}$ are  finite dimensional, which simplifies the problem of
diagonalization. Therefore the complete spectrum of the system can
be calculated through the diagonalization of the Hamiltonian in the
subspaces of all the unitary irreducible representations (UIR) of
$U^{B}(6)$, belonging to a given UIR of $Sp^{B}(12,R)$, which
further clarifies its role as a group of dynamical symmetry.

The most important application of the $U^{B}(6)\subset $
$Sp^{B}(12,R)$ limit of the theory is the possibility it affords of
describing both even and odd parity bands up to very high angular
momentum \cite{GGG}. In order to do this we first have to identify
the experimentally observed bands with the sequences of basis states
of the even $Sp^{B}(12,R)$ irrep (Table \ref{BasTab}). As we deal
with the symplectic extension we are able to consider all even
eigenvalues of the number of vector bosons $N$ with the
corresponding set of $T-$spins, which uniquely define the
$SU^{B}(3)$ irreps $(\lambda, \mu)$. The multiplicity index $K$
appearing in the final reduction to the $SO^{B}(3)$ is related to
the projection of $L$ on the body fixed frame and is used with the
parity ($\pi $)\ to label the different bands ($K^{\pi } $) in the
energy spectra of the nuclei. For the even-even nuclei we have
defined the parity of the states as $\pi_{core} =(-1)^{T}$
\cite{GGG}. This allowed us to describe both positive and negative
bands.

Further, we use the algebraic concept of \textquotedblleft
yrast\textquotedblright\ states, introduced in \cite{GGG}. According
to this concept we consider as yrast states the states with given
$L$ that minimize the energy with respect to the number of vector
bosons $N$ that build them. Thus the states of the ground state band
(GSB) were identified with the $SU^{B}(3)$ multiplets $(0,\mu )$
\cite{GGG}. In terms of $(N,T)$ this choice corresponds to $(N=2\mu
,T=0)$ and the sequence of states with different numbers of bosons
$N=0,4,8,\ldots $ and $T=0$ (and also $T_{0}=0$). The presented
mapping of the experimental states onto the $SU^{B}(3)$ basis
states, using the algebraic notion of yrast states, is a particular
case of the so called "stretched" states \cite{str}. The latter are
defined as the states with ($\lambda_{0}+2k,\mu_{0}$) or
($\lambda_{0},\mu_{0}+ k$), where $N_{i}=\lambda_{0}+2\mu_{0}$ and
$k=0,1,2,3, \ldots$. In the symplectic extension of the IVBM the
change of the number $k$, which is related in the applications to
the angular momentum $L$ of the states, gives rise to the collective
bands.

It was established \cite{GGD} that the correct placement of the
bands in the spectrum strongly depends on their bandheads'
configuration, and in particular, on the minimal or initial number
of bosons, $N = N_{i}$, from which they are built. The latter
determines the starting position of each excited band. In the
present application we take for $N_{i}$ the value at which the best
$\chi^2$ is obtained in the fitting procedure for the energies of
the considered excited band.

\begin{figure}[t]\centering
\includegraphics[width=110mm]{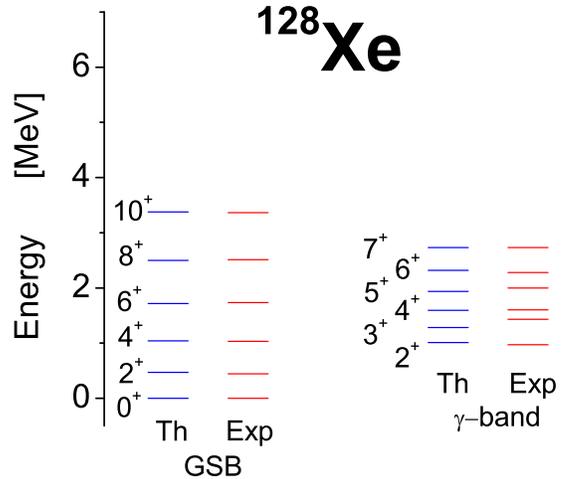}
\caption{(Color online) Comparison of the theoretical and
experimental energies for the ground and $\gamma$ bands of
$^{128}$Xe. The obtained values for the model parameters entering in
the core Hamiltonian (\ref{Hcore}) are: $a=0.08909$, $b=-0.00459$,
$\alpha_{3}=0.02471$, $\beta_{3}=0.03123$, and
$\alpha_{1}=-0.01989$.} \label{Xe128}
\end{figure}

The used Hamiltonian for the core nucleus is \cite{GGG}:
\begin{equation}
H_{B}=aN+bN^{2}+\alpha _{3}T^{2}+\beta _{3}L^{2}+\alpha
_{1}T_{0}^{2}, \label{Hcore}
\end{equation}
which is expressed in terms of the first and second order Casimir
operators from the unitary limit \cite{GGG} of the model (see the
boson part of chain (\ref{chain})). Taking into account  the
relation $N=\lambda+2\mu$ and $T=2\lambda$ between the quantum
numbers of the mutually complementary groups $SU^{B}(3)$ and
$U^{B}(2)$ in (\ref{chain}), it becomes obvious that $H_{core}$ is
diagonal in the basis
\begin{equation}
\mid [N]_{6};(\lambda,\mu);KLM;T_{0}\rangle \equiv \ \mid
(N,T);KLM;T_{0}\rangle, \label{bast}
\end{equation}
labeled by the quantum numbers of the subgroups of the chain
(\ref{chain}). Its eigenvalues are the energies of the basis states
of the boson representations of $ Sp(12,R)$:
\begin{eqnarray}
E((N,T),L,T_{0}) &=& aN + bN^{2} + \alpha_{3}T(T+1)  \nonumber \\
&+& \beta_{3}L(L+1)+\alpha_{1}T_{0}^{2}.  \label{Ecore}
\end{eqnarray}
We determine the values of the five phenomenological model
parameters $a, b, \alpha_{3}, \beta_{3}, \alpha_{1}$ by fitting the
energies of the ground and $\gamma-$ bands in $^{128}Xe$ nucleus to
the experimental data \cite{exp}, using a $\chi^{2}$ procedure. The
theoretical predictions are presented in Figure \ref{Xe128}. From
the figure one can see that the calculated energy levels  of both
ground state and $\gamma$ bands agree rather well with the observed
data.

\begin{figure}[h]\centering
\includegraphics[width=97mm]{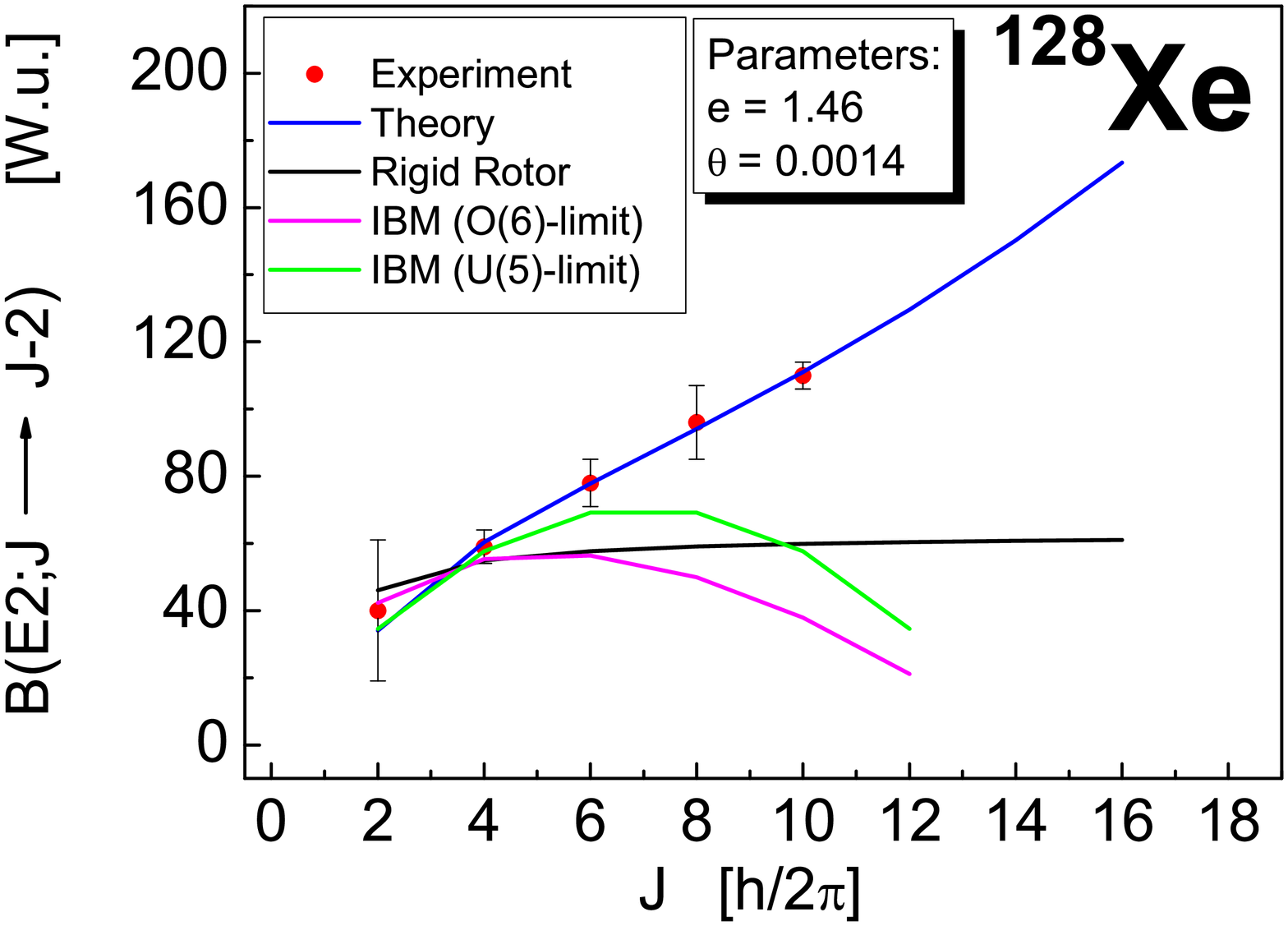}
\caption{(Color online) Comparison of the theoretical and
experimental values for the $B(E2)$ transition probabilities between
the states of GSB in $^{128}Xe$. The theoretical predictions of the
Rigid Rotor and IBM are shown as well. The values of the model
parameters are $e=1.46$ and $\theta= 0.0014$.} \label{E2-Xe128}
\end{figure}

Numerous IBM studies of even-even nuclei in the $A\sim 130$ mass
region have shown that these nuclei are well described by the $O(6)$
symmetry of the IBM, that in the classical limit corresponds to the
Wilets-Jean model of a $\gamma-$unstable rotor \cite{WJR}, and that
the accepted interpretation is that they are $\gamma-$soft. The core
nucleus $^{128}Xe$ has $R_{4/2}=2.33$ which is between the $U(5)$
and $O(6)$ values $R_{4/2}=2.00$ and $2.5$ respectively, which
reveals the transitional character of this $\gamma-$soft nucleus.
The value $R_{4/2}=2.33$ is close to the critical point value
$R_{4/2}=2.20$ for $E(5)$ symmetry, which has served as a ground for
some authors to consider $^{128}Xe$ as an $E(5)$ nucleus. In
\cite{z4}, the predictions of the $Z(4)$ model are compared to
existing experimental data for some nuclei, including $^{128}Xe$.
The reasonable agreement observed in \cite{z4} is in no
contradiction with the characterization of these nuclei as $O(6)$
nuclei, since it is known \cite{Casten} that $\gamma-$unstable
models (like $O(6)$) and $\gamma-$rigid models (like $Z(4)$) yield
similar predictions for most observables if $\gamma_{rms}$ of the
former equals $\gamma_{rigid}$ of the latter. In \cite{gCBS} a
$\gamma-$independent version of the confined beta-soft (CBS) rotor
model, in which the structure of $^{128}Xe$ was investigated, has
been formulated. That version, called $O(5)-$CBS, generalizes the
$E(5)$ solution near the critical point to a parametric solution for
the whole path between $E(5)$ and the $\beta-$rigidly deformed
$\gamma-$independent limit. The usage of all these models reveals
the transitional character of $^{128}Xe$ core nucleus. The
transitional properties of the latter can be clearly seen from its
electromagnetic properties. Indeed, one can see the characteristic
(almost "linear") $B(E2)$ behavior for the ground state band of
$^{128}Xe$ in Figure \ref{E2-Xe128}, where for comparison the
theoretical predictions of the rigid rotor and IBM in its $O(6)$ and
$U(5)$ limits are also shown. Such a behavior differs from the
typical parabolic-like (cut-off effect) $SU(3)$, $U(5)$ and $O(6)$
curves. The experimental data are taken from \cite{BE2Xe128}. The
obtained value for the quadrupole moment of the $2^{+}_{1}$ state of
GSB is $Q(2)=-0.15$ \emph{eb}. At this point we want to point out
that because of the mixing of different collective modes within the
framework of the symplectic IVBM \cite{GGG}, we are able to describe
even-even cores with various collective properties that need
different dynamical symmetries or their mixture in the IBM and other
algebraic approaches.

\section{Orthosymplectic extension}

In order to incorporate the intrinsic spin degrees of freedom into
the symplectic IVBM, we extend the dynamical algebra of
$Sp^{B}(12,R)$ to the orthosymplectic algebra of $OSp(2\Omega/12,R)$
\cite{OSE}. For this purpose we introduce a particle with spin
$j-$half-integer ($\Omega=2j+1$) and consider a simple
core-plus-particle picture. Thus, in addition to the boson
collective degrees of freedom we introduce creation and annihilation
operators $a_{m}^{\dag}$ and $a_{m}$ ($m=-j,\ldots,j$), which
satisfy the anticommutation relations:
$\{a_{m},a_{m'}^{\dag}\}=\delta _{mm'}$ and
$\{a_{m}^{\dag},a_{m'}^{\dag}\}=\{a_{m},a_{m'}\}=0$.

All bilinear combinations of \ $a_{m}^{+}$ and $a_{m'}$, namely
\begin{eqnarray}
f_{mm'} &=&a_{m}^{\dag}a_{m'}^{\dag}, \ \ m\neq m' \label{fs} \\
g_{mm'} &=&a_{m}a_{m'}, \ \ m\neq m'; \label{gs} \\
C_{mm'} &=&(a_{m}^{\dag}a_{m'}-a_{m'}a_{m}^{\dag})/2  \label{UN gen}
\end{eqnarray}%
generate the fermion-pair Lie algebra of $SO^{F}(2\Omega)$. Their
commutation relations are given in \cite{OSE}. The number preserving
operators (\ref{UN gen}) generate maximal compact subalgebra of
$SO^{F}(2\Omega)$, i.e. $U^{F}(\Omega)$.

\subsection{Fermion dynamical symmetries}

One can further construct a certain fermion dynamical symmetry, i.e.
the group-subgroup chain:
\begin{equation}
SO^{F}(2\Omega) \supset  G ' \supset  G '' \supset \ldots
.\label{FDS}
\end{equation}
In particular for one particle occupying a single level $j$ we are
interested in the following dynamical symmetry:
\begin{equation}
SO^{F}(2\Omega) \supset  Sp(2j+1) \supset SU^{F}(2),\label{opFDS}
\end{equation}
where $Sp(2j+1)$ is the compact symplectic group. The dynamical
symmetry (\ref{opFDS}) remains valid also for the case of two
particles occupying the same level $j$. In this case, the allowed
values of the quantum number $I$ of $SU^{F}(2)$ in (\ref{opFDS})
according to reduction rules are $I=0,2,\ldots,2j-1$ \cite{pair}.
For simplicity hereafter we will use just the reduction
$SO^{F}(2\Omega) \supset SU^{F}(2)$ and keep in mind the proper
content of the set of $I$ values for one and/or two particles cases,
respectively. For the $A \sim 130$ region where the doublet bands
are built on $\pi h_{11/2} \otimes \nu h_{11/2}$ configuration, the
two fermions occupy the same single particle level
$j_{1}=j_{2}=j=11/2$ with negative parity ($\pi_{sp}=-$) and the
fermion reduction chain (\ref{opFDS}) can be used.

\subsection{Dynamical Bose-Fermi symmetry. Dynamical supersymmetry}

Once the fermion dynamical symmetry is determined we proceed with
the construction of the Bose-Fermi symmetries. If a fermion is
coupled to a boson system having itself a dynamical symmetry (e.g.,
such as an IBM core), the full symmetry of the combined system is
$G^{B} \otimes G^{F}$. Bose-Fermi symmetries occur if at some point
the same group appears in both chains $G^{B} \otimes G^{F} \supset
G^{BF}$, i.e. the two subgroup chains merge into one.

The standard approach to supersymmetry in nuclei (dynamical
supersymmetry) is to embed the Bose-Fermi subgroup chain of $G^{B}
\otimes G^{F}$ into a larger supergroup G, i.e. $G \supset G^{B}
\otimes G^{F}$. Making use of the embedding $SU^{F}(2)\subset
SO^{F}(2\Omega)$ we use the orthosymplectic (supersymmetric)
extension of the IVBM which is defined through the chain \cite{OSE}:
\begin{equation}
\begin{tabular}{lllll}
$OSp(2\Omega/12,R)$ & $\supset $ & $SO^{F}(2\Omega)$ & $\otimes $ & $Sp^{B}(12,R)$ \\
&  &  &  & $\ \ \ \ \ \ \cup $ \\
&  & $\ \ \ \ \ \cup $ & $\otimes $ & \ $\ U^{B}(6)$ \\
&  &  &  & $\ \ \ \ \ N$ \\
&  &  &  & $\ \ \ \ \ \ \cup $ \\
&  & $SU^{F}(2)$ & $\otimes $ & $SU^{B}(3)\otimes U_{T}^{B}(2)$ \\
&  & $\ \ \ \ \ I$ &  & $(\lambda ,\mu )\Longleftrightarrow (N,T)~$ \\
&  & \multicolumn{1}{r}{$\searrow $} &  & $\ \ \ \ \ \ \cup $ \\
&  &  & $\otimes $ & $SO^{B}(3)\otimes SO^{B}_{T}(2)$ \\
&  &  &  & $~~~L\qquad \qquad T_{0}$ \\
&  &  & $\cup $ &  \\
&  & $Spin^{BF}(3)$ & $\supset $ & $Spin^{BF}(2),$ \\
&  & $~~~~~J$ &  & $~~~~~J_{0}$%
\end{tabular}
\label{chain}
\end{equation}%
where below the different subgroups the quantum numbers
characterizing their irreducible representations are given.
$Spin^{BF}(n)$ ($n=2,3$) denotes the universal covering group of
$SO(n)$.

At this point we want to stress that although the "coupling" of a
particle (or two-particle system) to the symplectic core is done at
the $SU(2)$ level, the present situation is not identical to that of
IBFM. In fact, in our approach due to the (ortho)symplectic
structures (allowing the change of number of phonon excitations $N$,
which in turn change the $SU(3)$ and $SO(3)$ content according to
the reduction rules) the core is no longer inert. In our application
the fermion angular momentum $I$ is algebraically added (subtracted)
to the changing core angular momentum $L$, i.e. the particle is
"dragged" around in the symplectic core. In some sense, in contrast
to the IBFM, the situation here is inverted: we have active boson
core and inert fermion part. Physically, this does \emph{not}
correspond to the "weak coupling limit" (as should be if $N$ was
fixed) between the core and the particle as it is in the case of
IBFM (on this level of coupling). In this way in the present
approach the nuclear dynamics is completely determined by the boson
degrees of freedom and the combined boson-fermion system is
essentially a new rotor with slightly different bulk properties,
such as moment of inertia, etc.

\section{The energy spectra of odd-mass and odd-odd nuclei}

We can label the basis states according to the chain (\ref{chain})
as:
\begin{eqnarray}
|~[N]_{6};(\lambda,\mu);KL;I;JJ_{0};T_{0}~\rangle \equiv \nonumber \\
|~[N]_{6};(N,T);KL;I;JJ_{0};T_{0}~\rangle ,  \label{Basis}
\end{eqnarray}
where $[N]_{6}-$the $U(6)$ labeling quantum number and
$(\lambda,\mu)-$the $ SU(3)$ quantum numbers characterize the core
excitations, $K$ is the multiplicity index in the reduction
$SU(3)\supset SO(3)$, $L$ is the core angular momentum, $I-$the
intrinsic spin of an odd particle (or the common spin of two fermion
particles for the case of odd-odd nuclei), $J,J_{0}$ are the total
(coupled boson-fermion) angular momentum and its third projection,
and $T$,$T_{0}$ are the $T-$spin and its third projection,
respectively.

The infinite set of basis states classified according to the
reduction chain (\ref{chain}) are schematically shown in Table
\ref{BasTab}. The fourth and fifth columns show the $SO^{B}(3)$
content of the $SU^{B}(3)$ group, given by the standard Elliott's
reduction rules \cite{Elliott}, while the next column gives the
possible values of the common angular momentum $J$, obtained by
coupling of the orbital momentum $L$ with the spin $I$. The latter
is vector coupling and hence all possible values of the total
angular momentum $J$ should be considered. For simplicity, only the
maximally aligned ($J=L+I$) and maximally antialigned ($J=L-I$)
states are illustrated in Table \ref{BasTab}.

\begin{center}
\begin{table}[h!]
\caption{Classification scheme of basis states (\protect\ref{Basis})
according the decompositions given by the chain
(\protect\ref{chain}).}
\smallskip \centering{\footnotesize \renewcommand{\arraystretch}{1.25}
\begin{tabular}{l|l|l|l|l|l}
\hline\hline $N$ & $T$ & $(\lambda ,\mu )$ & $K$ & $L$ & \quad \quad
\quad \quad \quad \quad \quad $J=L\pm I$ \\ \hline\hline $0$ & $0$ &
$(0,0)$ & $0$ & $0$ & $I$ \\ \hline\hline $2$ & $1$ & $(2,0)$ & $0$
& $0,2$ & $I;\ 2\pm I$ \\ \cline{2-6} & $0$ & $(0,1)$ & $0$ & $1$ &
$1\pm I$ \\ \hline\hline & $2$ & $(4,0)$ & $0$ & $0,2,4$ & $I;\ 2\pm
I;4\pm I$ \\ \cline{2-6}
$4$ & $1$ & $(2,1)$ & $1$ & $1,2,3$ & $1\pm I;\ 2\pm I;\ 3\pm I$ \\
\cline{2-6} & $0$ & $(0,2)$ & $0$ & $0,2$ & $I;\ 2\pm I$ \\
\hline\hline
& $3$ & $(6,0)$ & $0$ & $0,2,4,6$ & $I;\ 2\pm I;\ 4\pm I;6\pm I$ \\
\cline{2-6} & $2$ & $(4,1)$ & $1$ & $1,2,3,4,5$ &
\begin{tabular}{l}
$1\pm I;\ 2\pm I;\ 3\pm I;$ \\
$4\pm I;\ 5\pm I$%
\end{tabular}
\\ \cline{2-6}
$6$ & $1$ & $(2,2)$ & $2$ & $2,3,4$ & $2\pm I;\ 3\pm I;\ 4\pm I$ \\
\cline{2-6} &  &  & $0$ & $0,2$ & $I;\ 2\pm I$ \\ \cline{2-6} & $0$
& $(0,3)$ & $0$ & $1,3$ & $1\pm I;\ 3\pm I$ \\ \hline\hline
$\vdots $ & $\vdots $ & $\vdots $ & $\vdots $ & $\vdots $ & $\vdots $%
\end{tabular}
} \label{BasTab}
\end{table}
\end{center}
The basis (\ref{Basis}) can be considered as an angular-momentum
product of the orbital $\mid (N,T);KLM;T_{0} \ \rangle$ and spin
$\mid IM_{I} \ \rangle$ (or for the two particle case $\mid
I_{p},I_{n};IM_{I} \ \rangle$) wave functions. Then, if the parity
of the single particle is $\pi_{sp}$, the parity of the collective
states of the odd$-A$ nuclei will be $\pi=\pi_{core} \pi_{sp}$
\cite{OSE}. In analogy, one can write $\pi=\pi_{core} \pi_{sp}(1)
\pi_{sp}(2)$ for the case of odd-odd nuclei. Thus, the description
of the positive and/or negative parity bands requires only the
proper choice of the core band heads, on which the corresponding
single particle(s) is (are) coupled to, generating in this way the
different odd$-A$ (odd-odd) collective bands.

The Hamiltonian of the combined boson-fermion system can be written
as:
\begin{eqnarray}
H =H_{B}+H_{F}+H_{BF}, \label{Hamiltonian}
\end{eqnarray}
where the fermion degrees of freedom, coupled to the boson core, are
incorporated through the terms coming from the orthosymplectic
extension of the model:
\begin{eqnarray}
H_{F}+H_{BF}= \eta I^{2}+\gamma J^{2}+\zeta J_{0}^{2}. \label{HFBF}
\end{eqnarray}
The Hamiltonian (\ref{Hamiltonian}) is diagonal in the basis
(\ref{Basis}). Then its eigenvalues that yield the spectrum of the
odd-mass and odd-odd systems are:
\begin{eqnarray}
E(N;T,T_{0};L,I;J,J_{0})=aN+bN^{2} \nonumber \\
+\alpha _{3}T(T+1)+\beta _{3}L(L+1)+\alpha _{1}T_{0}^{2} \nonumber \\
+\eta I(I+1)+\gamma J(J+1)+\zeta J_{0}^{2}.\label{Energy}
\end{eqnarray}
The last term is needed only for the calculation of the energies of
the odd-mass neighboring nuclei.

Guided by the microscopic foundation of IBFM
\cite{IBFA1}-\cite{IBFA3}, where it is shown that the most important
term in the boson-fermion interaction is the dynamical (quadrupole)
one, we introduce in the Hamiltonian (\ref{Hamiltonian}) an
additional interaction between the core and the combined
two-particle system of quadrupole type:
\begin{eqnarray}
H_{int} = kQ_{B}\cdot Q_{F}, \label{Hint}
\end{eqnarray}
where $Q_{B}=\sqrt{6} \sum_{\alpha} A_{M}^{2}(\alpha,\alpha)$ and
$Q_{F}=Q_{\pi}+Q_{\nu}$ are the boson and fermion quadrupole
operators, respectively. The matrix element of $H_{int}$ between the
basis states (\ref{Basis}) can be written as
\begin{eqnarray}
&&\langle L'\tau',I';J | Q_{B} \cdot Q_{F} | L\tau,I; J
\rangle \nonumber \\
&&= (-1)^{J+L+I'} \left\{\begin{array}{c}
L \ J \ I \\
I' \ 2 \ L'%
\end{array} \right\}  \nonumber \\
&&\times\langle L'\tau'|| Q_{B} || L\tau \rangle \langle
j_{\pi}j_{\nu}I'|| Q_{F} || j_{\pi}j_{\nu}I \rangle, \label{MEQQ}
\end{eqnarray}
where $L'=L\mp2$, $I'=I\pm2$ and $\{_{I'2L}^{LJI}\}$ stands for $6j$
symbol. The labels $\tau$ and $\tau'$ denote the other quantum
numbers of the basis states in chain (\ref{chain}). The required
reduced matrix elements entering in (\ref{MEQQ}) are given in
Appendix B. {The important point here is that the boson ($L$) and
fermion ($I$) angular momenta constituting $J$ are changed by two
units in a way to preserve their sum always as $J'=J$. The
expectation value of $H_{int}$ will give a correction $\Delta
E=\Delta E(N_{0},L,I; n,j)$ to the energies (\ref{Energy}), but will
preserve the value of total (combined boson-fermion) angular
momentum $J$, characterizing each observed state of the yrast or
yrare band.

\section{Numerical results}

In our application, the most important point is the identification
of the experimentally observed states with a certain subset of basis
states from (ortho)symplectic extension of the model. In general,
except for the excited $\gamma-$band of the even-even nucleus
$^{128}Xe$ for which the stretched states of the first type
($\lambda-$changing) are used, the stretched states of the second
type ($\mu-$changing) are considered in all the calculations of the
collective states of the neighboring odd-mass and doubly odd nuclei.

In addition to the five parameters $a, b, \alpha_{3}, \beta_{3},
\alpha_{1}$ entering in Eq. (\ref{Energy}) which are fitted to the
energies of the even-even core nucleus, the number of adjustable
parameters needed for the complete description of the collective
spectra of both odd-A and odd-odd nuclei is four, namely $\gamma$,
$\zeta $, $\eta$ and $k$. The first two are evaluated by a fit to
the experimental data \cite{exp} of the lowest negative parity band
of the corresponding odd-A neighbor, while the last two are
introduced in the final step of the fitting procedure for the
odd-odd nucleus, respectively. Their numerical values are
$\gamma=0.00738$, $\zeta= 0.00986$, $\eta=0.11338$ and $k= 2.28115$.
Hence, as a result of the whole fitting procedure we are able to
describe simultaneously the energy spectra of the four neighboring
nuclei with the same set of parameters.

\begin{figure}[h]\centering
\includegraphics[width=100mm]{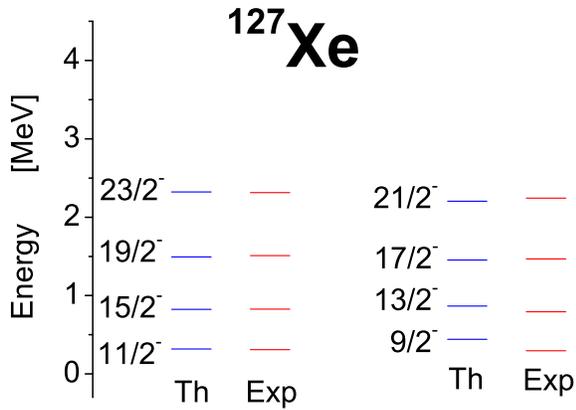}
\caption{(Color online) Comparison of the theoretical and
experimental energies for the lowest negative parity states built on
$h_{11/2}$ configuration in $^{127}Xe$.} \label{Xe127}
\end{figure}

\begin{figure}[h]\centering
\includegraphics[width=100mm]{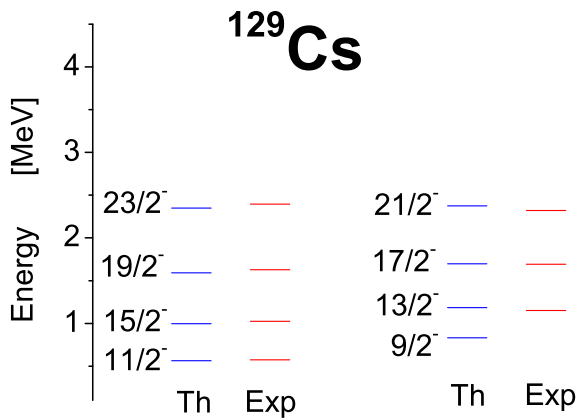}
\caption{(Color online) Comparison of the theoretical and
experimental energies for the lowest negative parity states built on
$h_{11/2}$ configuration in $^{129}Cs$.} \label{Cs129}
\end{figure}

The odd-A neighboring nucleus $^{127}Xe$ can be considered as a
neutron-hole coupled to the even-even core $^{128}Xe$. The low-lying
positive parity states of the GSB in odd-A neighbor are based on
positive parity proton and positive parity neutron configurations
($s_{\frac{1}{2}},d_{\frac{3}{2}},d_{\frac{5}{2}},g_{\frac{7}{2}}$),
whereas those of negative parity$-$ on $h_{11/2}$. Thus, we take
into account only the single particle orbit $j_{1}=11/2$. The
comparison between the experimental and calculated spectra for the
lowest negative parity band in $^{127}Xe$ is illustrated in Figure
\ref{Xe127}. One can see from the figure that the calculated energy
levels agree well with the experimental data. The comparison between
the experimental and calculated spectra (with the same set of
parameters) for the lowest negative parity structure in the
odd-proton neighbor $^{129}Cs$ is illustrated in Figure \ref{Cs129}.

\begin{figure}[h]\centering
\includegraphics[width=100mm]{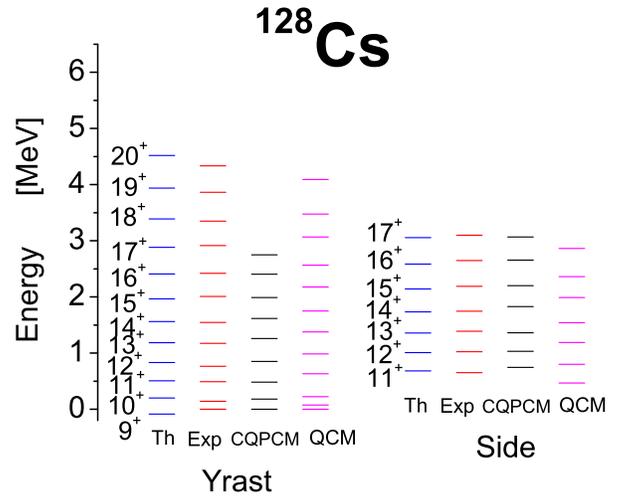}
\caption{(Color online) Comparison of the theoretical and
experimental energies for the yrast and side bands of $^{128}Cs$.
The theoretical predictions of the CQPCM are shown as well. The used
values of the model parameters are as follows: $a=0.08909$,
$b=-0.00459$, $\alpha_{3}=0.02471$, $\beta_{3}=0.03123$,
$\alpha_{1}=-0.01989$, $\gamma=0.00738$, $\zeta= 0.00986$,
$\eta=0.11338$ and $k= 2.28115$.} \label{Cs128}
\end{figure}

For the calculation of the odd-odd nuclei spectra a second particle
should be coupled to the core. In our calculations a consistent
procedure is employed which includes the analysis of the even-even
and odd-even neighbors of the nucleus under consideration. Thus, as
a first step an odd particle was coupled to the boson core
$^{128}Xe$ in order to obtain the spectra of the odd-mass neighbor
$^{127}Xe$. As a second step, we consider the addition of a second
particle on a single $j_{2}=11/2$ level to the boson-fermion system.
The level scheme presenting the doublet band structure in
$^{128}Cs$, where both the proton and neutron odd particles occupy
the same single particle level $j=j_{1}=j_{2}=11/2$, is shown in
Fig. \ref{Cs128}. For comparison, the CQPCM \cite{a6} and QCM
\cite{qcm2} results are also shown. In our model considerations, the
states of the doublet bands are mapped onto the stretched $SU(3)$
multiplets ($\lambda_{0},\mu_{0}$), ($\lambda_{0},\mu_{0}+1$),
($\lambda_{0},\mu_{0}+2$), \ldots, where the band head structures
for the yrast and side bands are determined by
($\lambda_{0}=0,\mu_{0}=11$) (or  $N_{0}=22$) and
($\lambda_{0}=8,\mu_{0}=9$) ($N_{0}=26$), respectively. From the
figure one can see that except the first ($9^{+}$) and last
($20^{+}$) positive states of yrast band, there is a very good
agreement between the theoretical predictions and experiment up to
very high angular momenta for both yrast and side bands, which
reveals the applicability of the used dynamical symmetry of the
model.

The similar level scheme of $^{128}Cs$ indicates that this nucleus
can be interpreted as having a chiral structure. However, the energy
splitting of these bands, connected with the extent of violation of
chiral symmetry \cite{breaking} is about $200$ keV. Although the
partner bands are not degenerated as should be in the ideal chiral
situation, this is the first indication that the partner bands in
$^{128}Cs$ might have properties closer to the expected features of
chiral bands. To investigate the structure of the doublet bands in a
certain nucleus, it is crucial to determine the $B(E2)$ and $B(M1)$
values which are very important for establishing the nature of these
bands. So, in the next section we consider the $E2$ and $M1$
transitions in the framework of the IVBFM.

\section{Electromagnetic transitions}

In this section we calculate the $B(E2)$ and $B(M1)$ transition
probabilities between the states of the partner bands and compare
them with their absolute values measured in the experiment
\cite{Cs128}.

The $E2$ transition operator between the states of the considered
band is defined as \cite{DB},\cite{TP}:
\begin{widetext}
\begin{equation}
T^{E2}=e\left[A_{(1,1)_{3}[0]_{2}\quad 00}^{ \lbrack 1-1]_{6}\quad
\quad 20} +\theta ([F\times F]_{(0,2)[0]_{2}\quad 00}^{\quad \lbrack
4]_{6}\quad \, \ 20}+[G\times G]_{(2,0)[0]_{2}\quad 00}^{\quad
\lbrack -4]_{6}\quad \,20})\right], \label{te2}
\end{equation}%
\end{widetext}
where the symbols $[ \ ]_{6}$, $( \ , \ )_{3}$ and $[ \ ]_{2}$
denote the corresponding $U(6)$, $SU(3)$ and $U(2)$ irreducible
representations, respectively. In (\ref{te2}) the following notation
$[A \times B]_{(\lambda,\mu)[2T]_{2} \ TT_{0}}^{\quad \lbrack
N]_{6}\quad \, \ LM}$ for the tensor coupling of two tensors A and B
to the respective final representations is also used. The first part
of (\ref{te2}) is a $SU(3)$ generator and actually changes only the
angular momentum with $\Delta L=2$, whereas the second term changes
both the number of bosons by $\Delta N=4$ and the angular momentum
by $\Delta L=2$. In (\ref{te2}) $e$ is the effective boson charge
fitted together with $\theta$ to the experimental data on the
transitions.

\begin{figure}[h]\centering
\includegraphics[width=98mm]{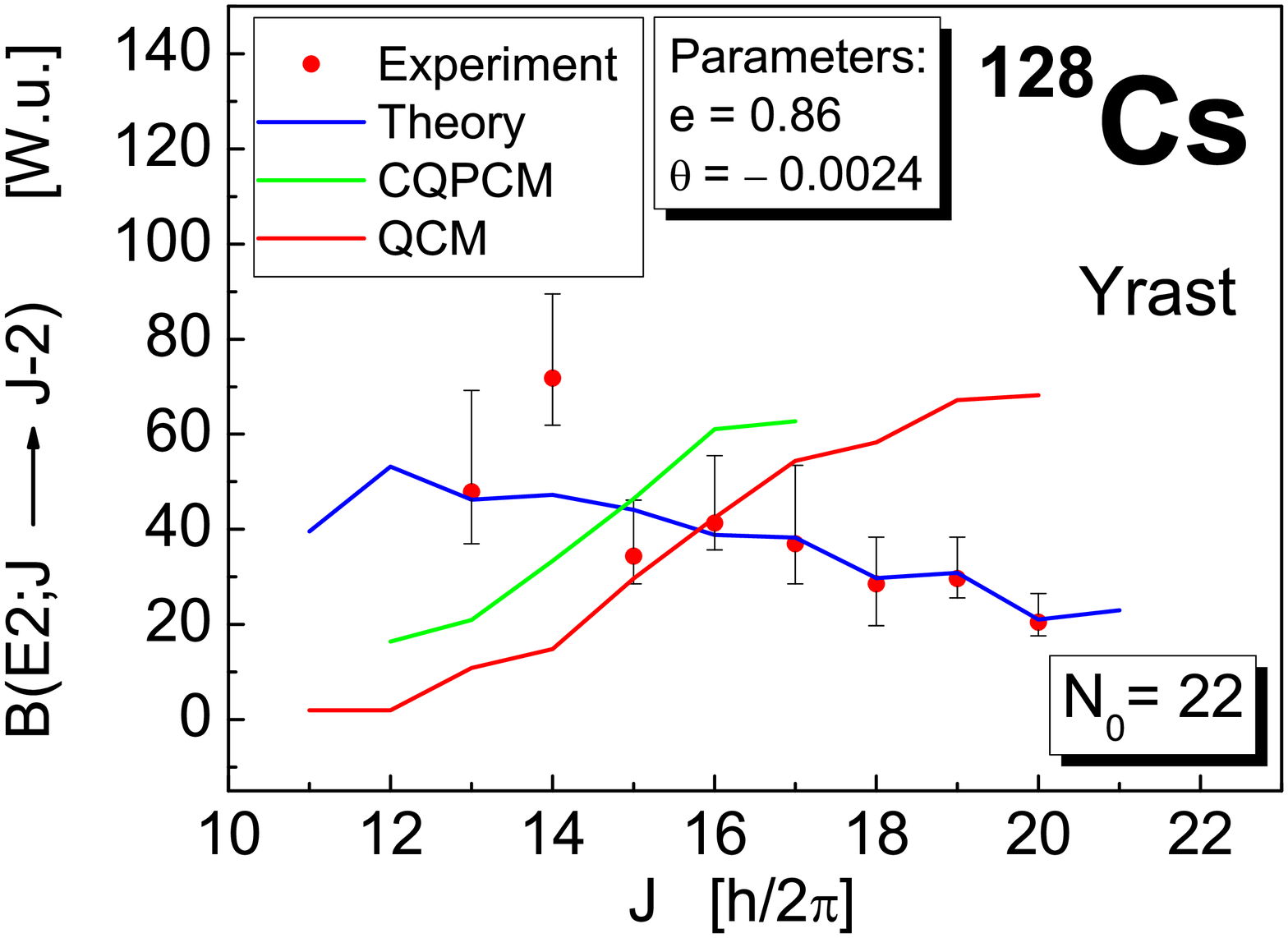}\hspace{1.mm}
\includegraphics[width=98mm]{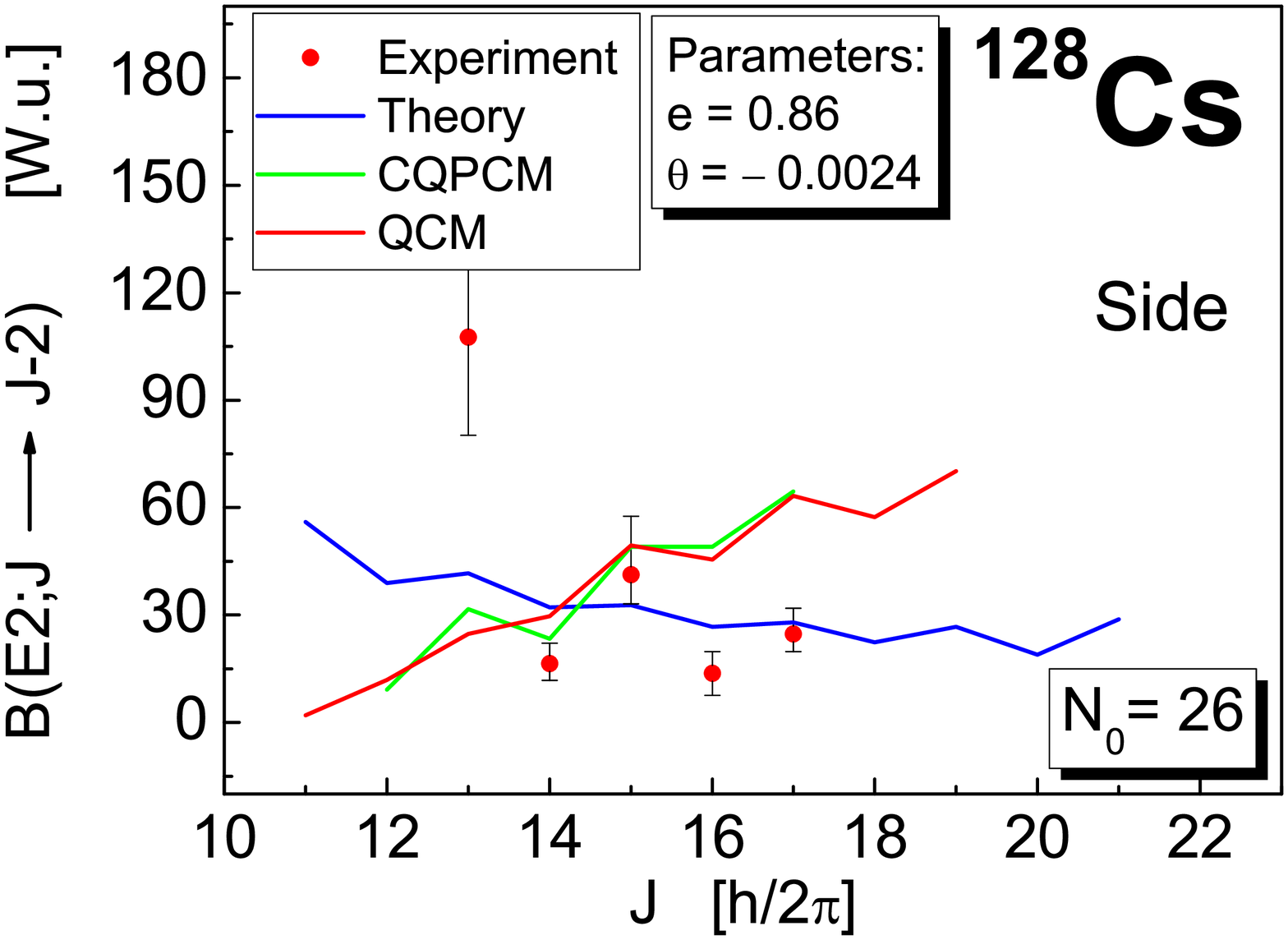}
\caption{(Color online) Comparison of the theoretical and
experimental values for the $B(E2)$ transition probabilities between
the states of yrast and side bands for $^{128}Cs$. The theoretical
predictions of the CQPCM and QCM are shown as well. The values of
the model parameters are $e=0.86$ and $\theta= -0.0024$.}
\label{CsE2o}
\end{figure}

The theoretical predictions for the $B(E2)$ values for $^{128}Cs$
are compared with the experimental data \cite{Cs128} and the CQPCM
\cite{Cs128} and QCM \cite{qcm2} results in Figure \ref{CsE2o}. The
used values for the two model parameters are $e=0.86$ and $\theta=
-0.0024$. In our approach, the odd-spin and even-spin members of the
yrast or side band form two $\Delta J=2$ $E2$ bands with the
sequences $J,J+2,J+4,\ldots$ and $J+1,J+3,\ldots$ which are built on
two different band head configurations having an intrinsic spin $I$
($=8$) and $I-2$ ($=6$), respectively. The  different two-particle
configurations $I$ are caused by the relative motion of the two
angular momenta of the proton and the neutron, which open and close
consequently ("scissors-like" motion) and thus changing $I$ by
$\Delta I=2$ in the following way: $I\rightarrow I-2 \rightarrow I$.
From the figure one can see the good overall reproduction of the
experimental values. The $B(E2)$ values in the yrast band are
$20-60$ percents larger than in side band.

In Figure \ref{QMCs128} we show the behavior of the quadrupole
moments $Q(J)$ as a function of the angular momentum $J$ for the
yrast and side bands in $^{128}Cs$.

\begin{figure}[h]\centering
\includegraphics[width=98mm,height=73mm]{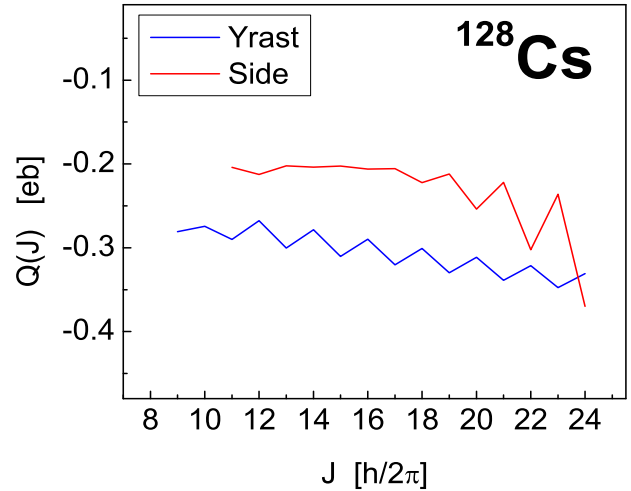}
\caption{(Color online) Theoretical values of the quadrupole moments
$Q(J)$ as a function of the angular momentum J for both yrast and
side bands in $^{128}Cs$.} \label{QMCs128}
\end{figure}

The structure of $M1$ transition operator between the states of the
yrast and yrare bands is defined as \cite{DB}:
\begin{widetext}
\begin{equation}
T^{M1 \quad(1)}_{\quad \quad M}=\sqrt{\frac{3}{4\pi}} \left[g
J_{M}^{(1)}+ g_{FG} (F_{(0,1)[0]_{2}\quad 00}^{\quad \lbrack
2]_{6}\quad \, \ 1M}+ G_{(1,0)[0]_{2}\quad 00}^{\quad \lbrack
-2]_{6}\quad \,1M}) \right]. \label{tm1}
\end{equation}
\end{widetext}
$J_{M}^{(1)}$ is the total boson-fermion angular momentum, i.e.
$J_{M}^{(1)}=L_{M}^{1}+ I_{M}^{(1)}$, where $L_{M}^{(1)}=-\sqrt{2}
\sum_{\alpha} A_{M}^{1}(\alpha,\alpha)$ and
$I_{M}^{(1)}=[a^{\dag}_{j}a_{j}]^{(1)}_{M}$. The obtained values of
effective g-factors are $g = 0.58 \mu_{N}$ and $g_{FG} = -1.41
\mu_{N}$.

\begin{figure}[h]\centering
\includegraphics[width=96mm]{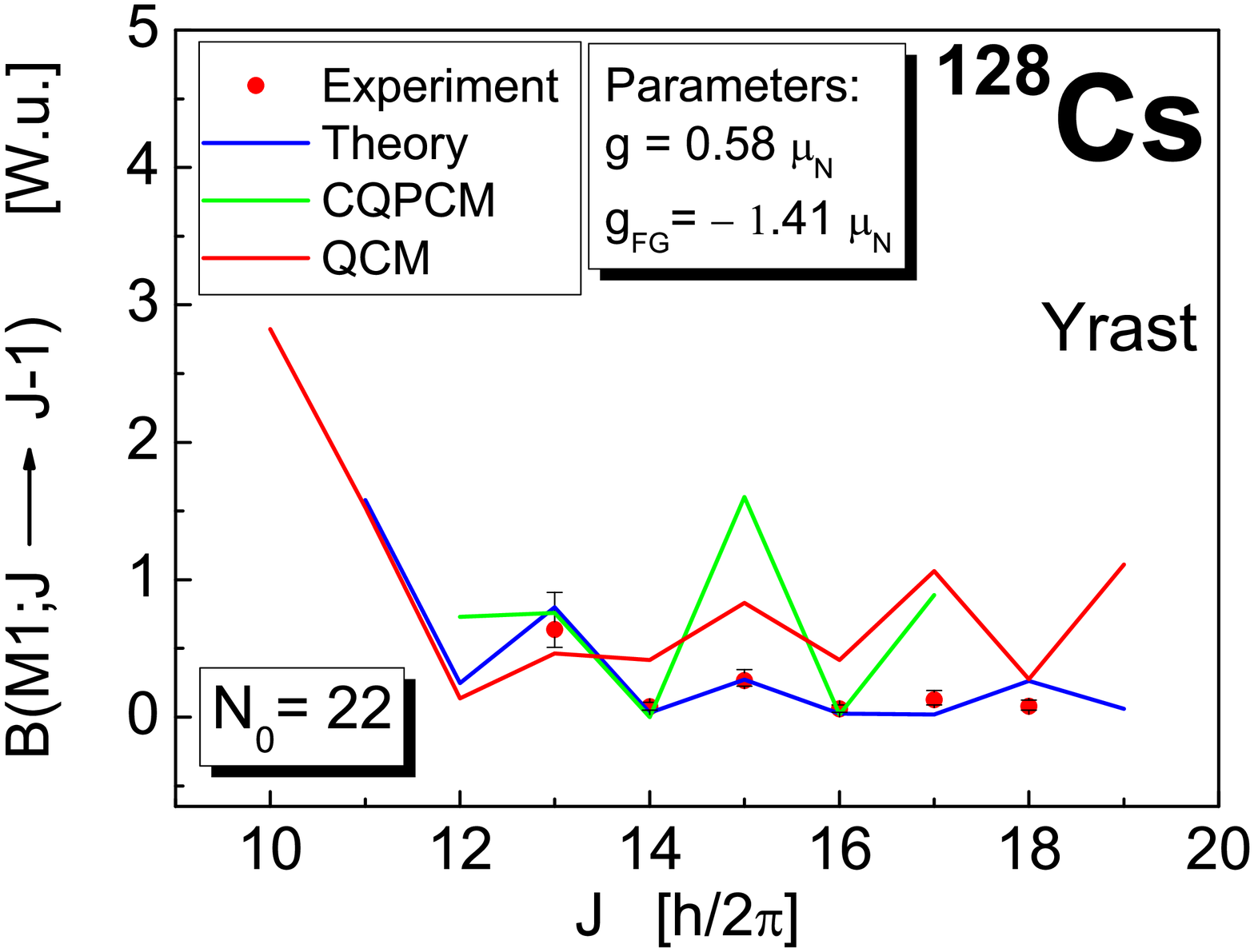}\hspace{1.mm}
\includegraphics[width=96mm]{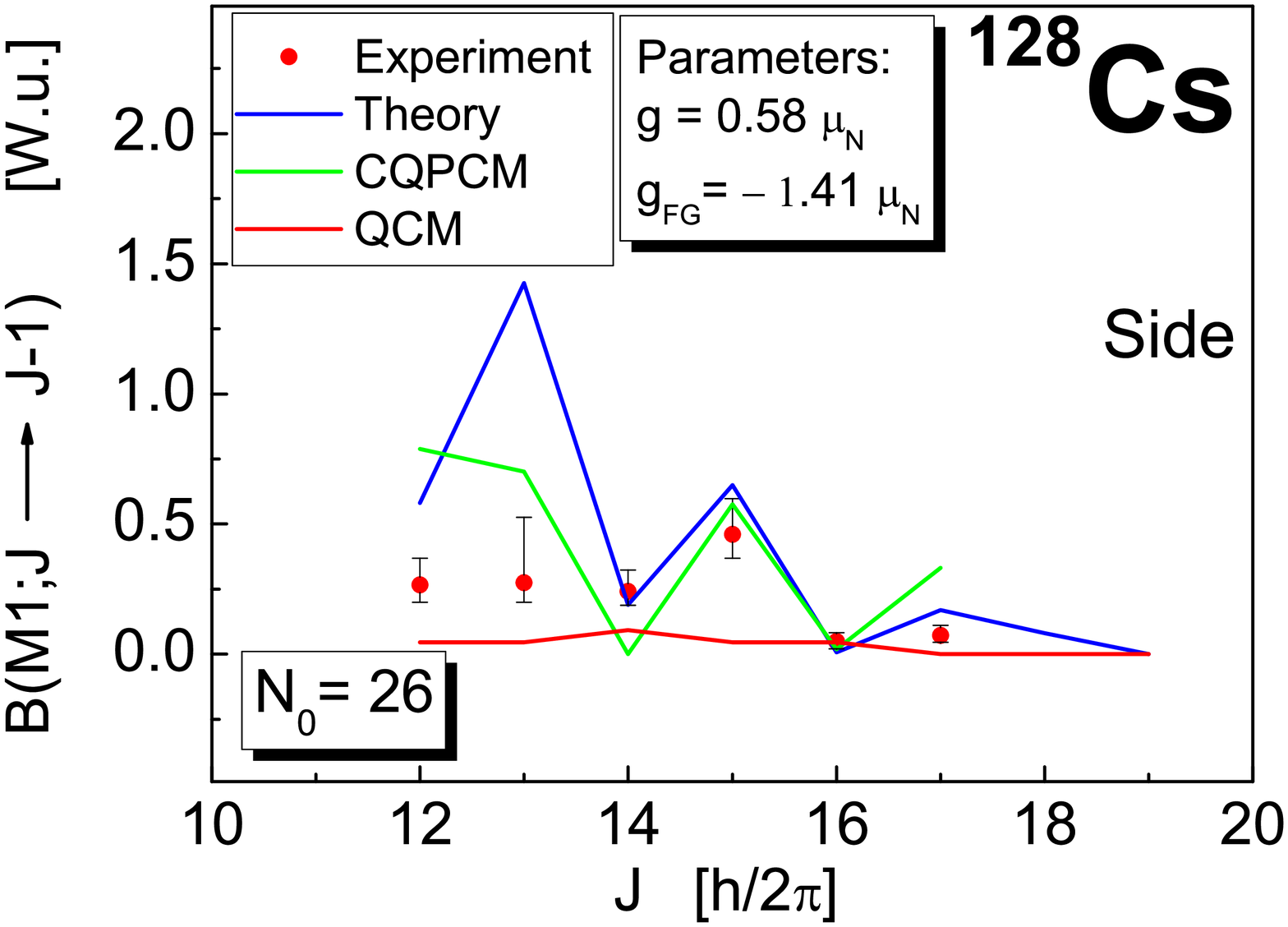}
\caption{(Color online) Comparison of the theoretical and
experimental values for the $B(M1)$ transition probabilities between
the states of yrast and side bands for $^{128}Cs$. The theoretical
predictions of the CQPCM and QCM are shown as well. The values of
the model parameters are $g = 0.58 \mu_{N}$ and $g_{FG} = -1.41
\mu_{N}$.} \label{CsM1o}
\end{figure}

The theoretical predictions for the intraband $B(M1)$ values for the
partner bands in $^{128}Cs$ are compared with the experimental data,
the CQPCM \cite{Cs128} and QCM \cite{qcm2} results in Figure
\ref{CsM1o}. The spin dependence of the reduced $M1$ transition
probabilities inside each of the bands show characteristic
staggering and, except the state $J=13$ of the side band, is
reproduced reasonably good. The $B(M1)$ values between the states of
both the yrast and side (less pronounced) bands for the transitions
from the odd-spin states to the even-spin states are larger than the
transitions from even-spin to odd-spin states. The strong $M1$
transitions connect the odd-spin states $(J+1)$ built on the
bandhead with the intrinsic spin $I$ to the even-spin states $(J)$
which possesses a bandhead configuration with $I-2$. The difference
of the two-particle structures for the two sequences with odd and
even values of $J$, caused by the scissors-like motion of the proton
and the neutron, is the reason that produces the staggering in the
$B(M1)$ behavior, which has the larger amplitude in the side band
than in the yrast one. The different amplitude of the $B(M1)$
staggering in the two bands  is a result of the different structure
of the used stretched states, which for the yrast and side bands are
determined by the $SU(3)$ multiplets $(0,\mu)$ and $(8,\mu)$,
respectively. The $B(M1)$ values in the yrast band are $5-30$ $\%$
larger than in side band. From the Fig. \ref{CsM1o} one can see that
the $B(M1)$ curves of QCM and IVBFM for the yrast band show very
similar behavior. At this point we want to point out that the
interpretation of the structure of doublet bands in the present work
is very similar to that given in Refs. \cite{ptsm}-\cite{qcm2}
especially the assignment of the states to the two $\Delta J=2$ $E2$
bands comprising the doublet bands and the corresponding quadrupole
transitions. Concerning the $M1$ transitions, the structure of the
$M1$ sequences in PTSM and QCM is rather different (especially at
the "bottom" of the bands) from that presented here.

In the picture of the chiral structure, the total angular momentum
is tilted with respect to the planes defined by the three principal
axes. This situation is realized when the angular momenta of the
valence proton, the valence neutron, and the triaxial core tend to
align with the short, long, and intermediate axes, respectively. It
indicates that the three considered angular momenta tend to be
perpendicular to each other. The deformation in the partner bands is
the same, as well as the structure of the corresponding states in
the partner bands.

Since the IVBFM is an algebraic model based on a set of algebraic
assumptions, the interpretation of its results in terms of space
orientation of the three angular momenta is not evident. First, as
we are interested in the dynamical symmetry $SO^{F}(2\Omega) \supset
Sp(\Omega) \supset SU^{F}(2)$, we approach the problem by
considering the simplest physical picture in which two particles
with the same intrinsic $j$ are coupled to an even-even nucleus.
This simplification that occurs when the fermion and boson degrees
of freedom are coupled on the level of the angular momenta (and
hence the neutron and proton states are not distinguished) weakens
the full fermion contribution to nuclear wave functions and
dynamics. If the two particles would be treated as different, a new
fermion dynamical chain should be considered in which both the spin
and isospin degrees of freedom are involved in the fermion sector,
that would introduce additional parameters in the energy formula and
making the identification of the bands parameter dependent.
Nevertheless, the result that the states of the partner bands are
built on proton-neutron configurations with angular momentum much
smaller than the maximal alignment, suggests that the condition of
their almost perpendicular orientation can be achieved. However,
this chiral-like  picture is disturbed by the scissors-like motion
of the two angular momenta of the neutron and the proton in the
two-particle system.

In the boson sector the adopted algebraic concept of "yrast" states
makes the geometrical interpretation less transparent. In our
approach the states of the yrast band are built on the ground state
band-like $SU(3)$ multiplets $(0,\mu)$ of the even-even core, while
those of the side band are built on the $SU(3)$ multiplets
$(8,\mu)$, which could suggest similar (both sets of $SU(3)$
multiplets are the stretched states of second type), but not equal
collective behavior of the two bands. This is reflected in the $E2$
transitions, possibly pointing to different deformations in partner
bands. The calculated quadrupole moments that are around 50 $\%$
smaller in the side than in the yrast band, strengthen the
conclusion that in IVBFM the two bands are built on different
deformations.

The results of our calculations show that the structure of
$^{128}Cs$ and its even-even and odd-even neighbors $^{128}Xe$,
$^{127}Xe$ and $^{129}Cs$ can be described in a consistent approach
based on the dynamical supersymmetry group $OSp(2\Omega /12, R)$.
This suggests that the appearance of doublet bands in certain
odd-odd nuclei could be a consequence of a larger symmetry
(supersymmetry), than those that arise in geometrical models. Over
the last few decades different supersymmetric extensions of the IBM
\cite{NSUSY} with its three boson dynamical symmetries for the core
were exploited, with Hamiltonians which exhibit dynamical
supersymmetries based on the compact supergroups of the type
$U(n/m)$. In this respect, the comparison of the theoretical
predictions of both compact and non-compact dynamical
supersymmetries is of particular interest. In the case of $^{128}Cs$
our calculations suggest that the symmetry is partially broken by
the additional interaction between the core and the two-fermion
system, but is still approximately realized.

\section{Conclusions}

In the present paper, the yrast and yrare states with the $\pi
h_{11/2} \otimes \nu h_{11/2}$ configuration in $^{128}Cs$ were
investigated in terms of the IVBFM. This allows for the proper
reproduction of the energies of these states up to very high angular
momenta in both bands. The even-even nucleus $^{128}Xe$ is used as a
core on which the collective excitations of the neighboring odd-mass
and odd-odd nuclei are built on. Thus, the spectra of odd-mass and
odd-odd nuclei arise as a result of the consequent and
self-consistent coupling of the fermion degrees of freedom to the
boson core. Therefore, according to our approach a purely collective
nature is assigned to the states of the partner bands.

The $B(E2)$ and $B(M1)$ transition probabilities between the states
of the yrast and side bands are calculated and compared with the
experimental data. A very good overall agreement of the theoretical
predictions with experiment, including the staggering patterns, is
obtained. The contribution of the symplectic term entering in the
corresponding transition operators turns out to be crucial for the
accurate reproduction of the experimental behavior. The observed
staggering in the $B(M1)$ (and $B(E2)$) values is reproduced within
our theoretical framework by introducing a quadrupole interaction
between the core and two-particle system, which produces such a
staggering effect in the transition strengths.

The consistent description of the structure of $^{128}Cs$ and its
even-even and odd-even neighbors $^{128}Xe$, $^{127}Xe$ and
$^{129}Cs$, suggest that the appearance of partner bands in certain
odd-odd nuclei could be a consequence of a larger symmetry
(supersymmetry), than those that arise in geometrical models,
pointing to the possible realization of non-compact dynamical
supersymmetries in heavy nuclei.

\section*{Acknowledgments}

We thank Alberto Ventura and Ana Georgieva for discussions. This
work was supported by the Bulgarian National Foundation for
scientific research under Grant Number DID-$02/16/17.12.2009$. H. G.
G. acknowledges also the support from the European Operational
programm HRD through contract $BG051PO001/07/3.3-02$ with the
Bulgarian Ministry of Education. \\

\appendix \textbf{Appendix A. Two-particle state of neutron and
proton} \setcounter{section}{1}

We consider a two-particle system of one neutron and one proton
particle in the same orbital $j$, where $j$ symbolically represents
the quantum numbers $(n,l,j)$. The wave function characterized by
the total spin $I$ and its projection $M_{I}$ is written as
\begin{eqnarray}
&&\mid jj;IM_{I} \rangle =\sum_{M_{1}M_{2}} (jM_{1}jM_{2}|IM) \mid
jM_{1} \rangle_{p} \mid jM_{2} \rangle_{n} \nonumber \\
&&=\left[\mid j \rangle_{p} \otimes \mid j \rangle_{n} \right]^{I}
_{M_{I}}. \label{tps}
\end{eqnarray}
where $\mid jM \ \rangle_{\tau}$ ($\tau=p$ or $n$) denotes a
single-particle state and $(j,m)$ represents a set of quantum
numbers necessary to specify the state $(n,l,j,m)$. We adopt
$j=11/2$, $l=j-1/2=5$, and $n=0$ to represent the intruder orbital
$0h_{11/2}$ in the $50-82$ major shell.

\appendix \textbf{Appendix B. Reduced matrix elements}
\setcounter{section}{2}

The reduced matrix element $\langle L'\tau'|| Q_{B} || L\tau
\rangle$ of the boson quadrupole operator $Q_{B}=\
A_{(1,1)_{3}[0]_{2} \ 00 }^{ \lbrack 1-1]_{6} \quad 2M}$ is given
by\cite{TP}:
\begin{equation}
\begin{tabular}{l}
$\langle \lbrack N],(\lambda ^{\prime },\mu ^{\prime });K^{\prime
}L^{\prime};T^{\prime }T_{0}^{\prime }|| A_{(1,1)[0]_{2} \
00}^{[1,-1]_{6}\
\ lm}||[N],(\lambda ,\mu );KL;TT_{0}\rangle $ \\
\\
$=\delta _{TT^{\prime }}\delta _{T_{0}T_{0}^{\prime }}\delta
_{\lambda \lambda ^{\prime }}\delta _{\mu \mu ^{\prime }}\sum_{\rho
=1,2}C_{K(L)\ \ \ k(l)\ \ K^{\prime }(L^{\prime })}^{(\lambda ,\mu
)\ \ (1,1)\ \
\rho (\lambda ^{\prime },\mu ^{\prime })}$ \\
\\
$\times \langle \lbrack N],(\lambda ^{\prime },\mu ^{\prime })|||\
A_{(1,1)[0]_{2}}^{[1,-1]_{6}\ }\
|||[N],(\lambda ,\mu )\rangle$, \\
\label{DRMEA}
\end{tabular}
\end{equation}
where $C_{KL\quad k(l)_{3}\quad K^{\prime }L^{\prime }}^{(\lambda
,\mu )\quad[\lambda ]_{3}\quad (\lambda ^{\prime },\mu ^{\prime })}$
is the reduced $SU(3)$ Clebsch-Gordan coefficient and the reduced
triple-barred matrix element for $\rho=1$ is
\begin{equation}
\langle \lbrack N],(\lambda ,\mu )||| \ A_{(1,1)_{3}[0]_{2} \ }^{
\lbrack 1-1]_{6}}|||[N],(\lambda ,\mu )\rangle _{1} = \left\{
\begin{array}{c}
g_{\lambda \mu }, \ \ \mu =0 \\
-g_{\lambda \mu }, \  \mu \neq 0%
\end{array}
\right. \label{TRMEA}
\end{equation}
where
\begin{equation}
g_{\lambda \mu }=2\left(\frac{\lambda ^{2}+\mu ^{2}+\lambda \mu
+3\lambda +3\mu }{3}\right)^{1/2}.
\end{equation}
For $\rho=2$ we have
\begin{equation}
\langle \lbrack N],(\lambda ,\mu )|||A_{[210]_{3}[0]_{2}\quad
}^{\quad \lbrack 1-1]_{6}}|||[N],(\lambda ,\mu )\rangle _{2}=0 .
\end{equation}
The reduced matrix element of a quadrupole operator for the
two-particle state is written as
\begin{eqnarray}
&&\langle jjI'|| Q_{\tau} || jjI \rangle =
(-1)^{2j+I}\sqrt{(2I+1)(2I'+1)} \nonumber \\
&&\times \left\{\begin{array}{c}
I' \ 2 \ I \\
j  \ j  \ j
\end{array} \right\} \langle j|| Q_{\tau} || j \rangle,  \label{RMEF}
\end{eqnarray}
where the reduced matrix element $\langle j|| Q_{\tau} || j \rangle$
is given by
\begin{eqnarray}
\langle j|| Q_{\tau} || j \rangle=\langle nl| r^{2} |nl \rangle
\langle j|| Y^{2} || j \rangle,
\end{eqnarray}
where $\tau=\pi,\nu$ and
\begin{eqnarray}
&&\langle nl| r^{2} |nl \rangle = (2n+l+3/2)=j+1,  \label{RMER}
\\  \nonumber \\
&&\langle j|| Y^{2}|| j \rangle  \nonumber \\
&&=(-1)^{j+1/2}\sqrt{\frac{5(2j+1)}{4\pi}} \left(\begin{array}{c}
j  \ \ \ 2  \ \ j \\
\frac{1}{2} \ \ 0  \ -\frac{1}{2}
\end{array} \right).
\label{RMEY}
\end{eqnarray}
Here, again we take $l=j-1/2$ and $n=0$. \\


\begin{thebibliography}{99}

\bibitem{a1} K. Starosta et al., Phys. Rev. Lett. \textbf{86}, 971
(2001).

\bibitem{a2} A. A. Hecht et al., Phys. Rev. \textbf{C 63}, 051302(R)
(2001).

\bibitem{a3} T. Koike et al., Phys. Rev. \textbf{C 63}, 061304
(2001).

\bibitem{a4} D. J. Hartley et al., Phys. Rev. \textbf{C 64}, 031304(R)
(2001).

\bibitem{a5} R. A. Bark et al., Nucl. Phys. \textbf{A 691}, 577
(2001).

\bibitem{a6} T. Koike et al., Phys. Rev. \textbf{C 67}, 044319 (2003).

\bibitem{cphcm} K. Starosta et al., Phys. Rev. \textbf{C 65}, 044328
(2002).

\bibitem{a8} G. Rainovski et al., Phys. Rev. \textbf{C 68}, 024318
(2003).

\bibitem{FM} S. Frauendorf, J. Meng, Nucl. Phys. \textbf{A 617}, 131 (1997).

\bibitem{tac3} V.I. Dimitrov, S. Frauendorf, F. D¨onau, Phys. Rev. Lett.
\textbf{84}, 5732 (2000).

\bibitem{cqpcm2} A.J. Simons et. al., J. Phys. G \textbf{3}1, 541 (2005).

\bibitem{prm3} T. Koike, K. Starosta, I. Hamamoto, Phys. Rev. Lett.
\textbf{93}, 172502 (2004).

\bibitem{tqptrm} I. Ragnarsson and P. Semmes, Hyperfine Interact. \textbf{43},
423 (1988).

\bibitem{soft} Ch. Droste et al., Eur. Phys. J. A \textbf{42}, 79 (2009).

\bibitem{contra1} S. Brant, D. Vretenar, and A. Ventura, Phys. Rev. \textbf{C 69},
017304 (2004).

\bibitem{contra2} E. Grodner, J. Srebrny, Ch. Droste, T. Morek, A. Pasternak, J.
Kownacki, Int. J. Mod. Phys. E \textbf{13}, 243 (2004).

\bibitem{contra3} D. Tonev et. al., Phys. Rev. Lett. \textbf{96}, 052501
(2006).

\bibitem{contra4} C.M. Petrache, G.B. Hagemann, I. Hamamoto, K. Starosta, Phys.
Rev. Lett. \textbf{96}, 112502 (2006).

\bibitem{breaking} K. Starosta, T. Koike, C. J. Chiara, D. B. Fossan, and D. R. Lafosse,
Nucl. Phys. \textbf{A682}, 375c (2001).

\bibitem{M1} C. Vaman, D. B. Fossan, T. Koike, K. Starosta, I. Y. Lee, and
A. O. Macchiavelli, Phys. Rev. Lett. \textbf{92}, 032501 (2004).

\bibitem{chivib} S. Mukhopadhyay et al., Phys. Rev. Lett. \textbf{99}, 172501 (2007).

\bibitem{M1stag} B. Qi, S. Q. Zhang, S. Y. Wang, J. M. Yao, and J. Meng,
Phys. Rev. C \textbf{79}, 041302(R) (2009).

\bibitem{ptsm} K. Higashiyama, N. Yoshinaga, Prog. Theor. Phys.
\textbf{113}, 1139 (2005); K. Higashiyama, N. Yoshinaga, K. Tanabe,
Phys. Rev. \textbf{C 72}, 024315 (2005); N. Yoshinaga, K.
Higashiyama, J. Phys. G \textbf{31}, S1455 (2005).

\bibitem{qcm1} N. Yoshinaga, K. Higashiyama, Eur. Phys. J. A \textbf{30},
343 (2006); \textbf{31}, 395 (2007).

\bibitem{qcm2} K. Higashiyama and N. Yoshinaga, Eur. Phys. J. A \textbf{33}, 355
(2007).

\bibitem{IBFFM} V. Paar, \textit{Capture Gamma-Ray Spectroscopy and Related Topics}, edited by
S. Raman, AIP Conf. Proc. No. 125 (AIP, New York, 1985), p. 70; S.
Brant, V. Paar, and D. Vretenar, Z. Phys. \textbf{A319}, 355 (1984);
V. Paar, D. K. Sunko, and D. Vretenar, Z. Phys. \textbf{A327}, 291
(1987); S. Brant and V. Paar, Z. Phys. \textbf{A329}, 151 (1988).

\bibitem{OSE} H. G. Ganev, J. Phys. G: Nucl. Part.
Phys. \textbf{35}, 125101 (2008).

\bibitem{DB} H. G. Ganev, A. I. Georgieva, S. Brant, and A. Ventura,
Phys. Rev. C \textbf{79}, 044322 (2009); H. G. Ganev, A. I.
Georgieva, S. Brant, and A. Ventura, Proceedings of $28^{th}$
International Workshop on Nuclear Theory, (June 21-27, 2009, Rila
Mountains, Bulgaria), ed. S. Dimitrova, Printed by BM Trade Ltd.,
Sofia, Bulgaria (2009) Rila Mountains,  pp. 177.

\bibitem{Cs128} E. Grodner et. al., Phys. Rev. Lett. \textbf{97}, 172501 (2006).

\bibitem{Cs126} X. Li et al., Chin. Phys. Lett.
\textbf{19}, 1779 (2002); S. Wang et al., Phys. Rev. C \textbf{74},
017302 (2006); S. Y. Wang, S. Q. Zhang, B. Qi, and J. Meng, Phys.
Rev. \textbf{C 75}, 024309 (2007).

\bibitem{Cs122} Yong-Nam U et. al., J. Phys. G: Nucl. Part. Phys. \textbf{31},
B1–B6 (2005).

\bibitem{GGG} H. Ganev, V. P. Garistov, and A. I. Georgieva,
Phys. Rev. C \textbf{69}, 014305 (2004).

\bibitem{str} D. J. Rowe, Rep. Prog. Phys. \textbf{48}, 1419
(1985).

\bibitem{GGD} H. G. Ganev, V. P. Garistov, A. I. Georgieva, and J. P. Draayer, Phys. Rev.
\textbf{C 70}, 054317 (2004).

\bibitem{exp} Evaluated Nuclear Structure Data File (ENSDF),
http://ie.lbl.gov/databases/ensdfserve.html; Level Retrieval
Parameters, http://iaeand.iaea.or.at/nudat/levform.html.

\bibitem{WJR} L. Wilets and M. Jean, Phys. Rev.
\textbf{102}, 788 (1956).

\bibitem{z4} D. Bonatsos et al., Rom. Journ. Phys., Vol. \textbf{52},
Nos. 8–10, P. 779–787, Bucharest, 2007.

\bibitem{Casten} R. F. Casten, \emph{Nuclear Structure from a Simple Perspective}, Oxford
University Press, Oxford, 1990.

\bibitem{gCBS} D. Bonatsos et al., Phys. Rev.
\textbf{C 74}, 044306 (2006).

\bibitem{BE2Xe128} Nuclear Levels and Gammas Search,
http://www.nndc.bnl.gov/.

\bibitem{pair} B. G. Wybourne,
\textit{Classical Groups for Physicists} (Wiley, New York, 1974).

\bibitem{Elliott} J. P. Elliott, Proc. R. Soc. \textbf{A245}, 128, 562 (1958).

\bibitem{IBFA1} I. Talmi, in \emph{"Interacting Bose-Fermi Systems in
Nuclei"} (F. Iachello, Ed.), p. 329, Plenum, New York, 1981.

\bibitem{IBFA2} O. Scholten and A. E. L. Dieperink, in \emph{"Interacting Bose-Fermi Systems in
Nuclei"} (F. Iachello, Ed.), p. 343, Plenum, New York, 1981.

\bibitem{IBFA3} O. Scholten, Prog. Part. Nucl. Phys.
\textbf{14}, 189 (1985).

\bibitem{TP} H. G. Ganev and A. I. Georgieva, Phys. Rev.
\textbf{C 76}, 054322 (2007).

\bibitem{NSUSY} F. Iachello, Phys. Rev Lett. \textbf{44}, 772 (1980); A. B.
Balantekin, I. Bars, R. Bijker, and F. Iachello, Phys. Rev.
\textbf{C 27}, 1761 (1983); P. Van Isacker, J. Jolie, K. Heyde, and
A. Frank, Phys. Rev Lett. \textbf{54}, 653 (1985); A. Metz et. al.,
Phys. Rev Lett. \textbf{83}, 1542 (1999); J. Barea, R. Bijker, A.
Frank, and G. Loyola, Phys. Rev. \textbf{C 64}, 064313 (2001); J.
Barea, R. Bijker, and A. Frank, J. Phys. A: Math. Gen. \textbf{37},
10251 (2004); J. Barea, R. Bijker, and A. Frank, Phys. Rev Lett.
\textbf{94}, 152501 (2005).



\end{thebibliography}
\end{document}